\documentclass{aa} \usepackage{times,graphicx}

\begin{document}

   \thesaurus{06           
     (13.07.1;)
              (13.09.2;)}  

            \title{Detection of the near-infrared counterpart of GRB 971214
              3.2 hours after the gamma-ray event.\thanks{Based on
                observations carried out at the German-Spanish Astronomical
                Center, Calar-Alto, operated by the Max-Planck-Institut
                f\"ur Astronomie, Heidelberg, jointly with the Spanish
                National Commission for Astronomy.}}

 
\author{
J. Gorosabel \inst{1} \and 
A. J. Castro-Tirado \inst{1,2}  \and 
C. J. Willott \inst{3}  \and 
H. Hippelein \inst{4}   \and
J. Greiner \inst{5}     \and 
A. Shlyapnikov \inst{6} \and 
S. Guziy \inst{7}       \and 
E. Costa \inst{8}       \and 
M. Feroci \inst{8}      \and 
F. Frontera \inst{9}    \and 
L. Nicastro \inst{10}    \and 
E. Palazzi \inst{11} }

            \offprints{J. Gorosabel (jgu@laeff.esa.es)}

            \institute{ Laboratorio de Astrof\'{\i}sica Espacial y
              F\'{\i}sica Fundamental (LAEFF-INTA), P.O. Box 50727, E-28080
              Madrid, Spain.  \and Instituto de Astrof\'{\i}sica de
              Andaluc\'{\i}a (IAA-CSIC), P.O. Box 03004, E-18080 Granada,
              Spain.  \and Astrophysics, Department of Physics, Keble Road,
              Oxford, OX1 3RH, U.K. \and Max-Planck-Institut f\"ur
              Astronomie, K\"onigstuhl, 69117 Heidelberg, Germany\and
              Astrophysikalisches Institut Potsdam, 14482 Potsdam, Germany.
              \and Space Research Group, Nikolaev Astronomical Observatory,
              Astronomical Observatory, Observatornaja, 327030 Nikolaev,
              Ukraine.  \and Nikolaev Astronomical Observatory, 327000
              Nikolaev, Ukraine.  \and Istituto di Astrofisica Spaziale,
              CNR, Frascati, Italy.  \and Dipartamento di Fisica,
              Universit\`a di Ferrara, 44100 Ferrara, Italy.  \and Istituto
              Astrofisica Spaziale, CNR, Via Fosso del Cavaliere, I-00131
              Roma, Italy.  \and Istituto Tecnologie e Studio Radiazioni
              Extraterrestri, CNR, Via P. Gobetti, 40129 Bologna, Italy.}

\date{Received date; accepted date}

\titlerunning{Near-IR detection of GRB 971214.}

\authorrunning{Gorosabel et al.}

            \maketitle

   \begin{abstract}
     We report the detection of the GRB 971214 counterpart in the
     near-infrared by means of two images in the K$^{\prime}$-band taken at
     Calar Alto only $\sim$3.5 and $\sim$5 hours after the gamma-ray event.
     We detect the transient at K$^{\prime}$=$18.03\pm0.18$ and
     K$^{\prime}$=$18.00\pm0.22$ respectively. Our data seem to indicate
     the existence of a plateau with duration $ 1.5 \leq T \le 6.7 $ hours
     (between 3.5 and 10.2 hours after the high-energy event). Moreover the
     power-law decline should be steeper than the one given by the index
     $\alpha_{\rm K^{\prime}}=0.45$. There is also a change in the slope of
     the broad-band spectrum at some wavelength between the J and
     K$^{\prime}$ bands (possibly around the H-band).


   \end{abstract}

%

\section{Introduction}
After 31 years since the discovery of gamma-ray bursts (GRBs), the origin
of such brief gamma-ray flashes remains unknown. The observed isotropy of
GRBs in the sky, could only be explained by theoretical models where GRBs
originate either in a extended halo around the Galaxy or arise from sources
at cosmological distances. Before the launch of the BeppoSAX satellite, the
poor localization capability of the GRB detectors made the searches at
other wavelengths unfruitful. 

The breakthrough took place in 1997 when the first X-ray afterglows were
observed by the BeppoSAX, RXTE, ROSAT and ASCA satellites (Costa et al.
1997, Heise et al.  1997a, Marshall et al. 1997, Greiner et al. 1998,
Murakami 1998).  They were able to localize the fading X-ray emission that
followed the more energetic gamma-ray photons once the GRB event had ended.
This emission (the afterglow) extends to longer wavelengths, and the good
accuracy in the position determination by BeppoSAX (typically $1^{\prime}$
radius error boxes) has led to the discovery of the first optical
counterparts for GRB 970228 (van Paradijs et al. 1997, Guarnieri et al.
1997), and GRB 970508 (Bond 1997, Djorgovski et al.  1997, Castro-Tirado et
al.  1998), greatly improving our understanding of these puzzling sources.
The measurement of the redshift for the GRB 970508 optical counterpart
(Metzger et al.  1997) has established that one GRB, maybe all, lie at
cosmological distances.

GRB 971214 is the third GRB with a known optical counterpart. It was
detected by the BeppoSAX Gamma-ray Burst Monitor (GBM, Frontera et al.
1997) on Dec 14.97 1997, as a 25 s long-structured gamma-ray burst.
Simultaneous to the detection of the GBM, the Wide Field Cameras (WFC,
Jager et al.  1997) on board BeppoSAX provided an accurate position (a
$3^{\prime}.9$ radius error box at a 3$\sigma$ confidence level, Heise et
al. 1997b) that allowed deep optical, infrared and radio observations. The
position was also consistent with the one given by the all-sky monitor on
RXTE (Bradt et al. 1993) and by the BATSE/Ulysses annulus (Kippen et al.
1997). When BeppoSAX pointed its Narrow-Field Instruments (NFI) to the GRB
position, on Dec 15.25 ($\sim$ 6.5 hours after the burst), a previously
unknown variable X-ray source was found inside the WFC error box (Antonelli
et al.  1997) which was identified as the X-ray afterglow of GRB 971214.
Soon after, Halpern et al.  (1997) reported the presence of a fading object
inside the WFC GRB error box, based on two I-band images separated 24
hours. The object was afterwards confirmed as the counterpart of GRB 971214
by means of additional observations at other wavelengths: R-band (Castander
et al.  1997, Diercks et al.  1998), I-band (Rhoads 1997) and J-band
(Tanvir et al.  1997). No detections in the K-band were reported in the
literature, although observations performed on Dec 15.54 imposed an upper
limit of K $>$ 18.5 (Garcia et al.  1997).  As it will be explained later,
this upper limit will be used to constraint the power-law index
$\alpha_{\rm K^{\prime}}$ and the position of the possible maximum of the
light curve.

We report here the detection of the GRB 971214 counterpart in the near
infrared (IR) by means of two K$^{\prime}$-band images taken at Calar Alto
on Dec 15.12 and 15.18 (mean observing time, only $\sim$3.5 and $\sim$5
hours after the gamma-ray event). The second image is almost simultaneous
to the beginning of the observations performed by the BeppoSAX narrow-field
instruments.  We discuss whether our observations are in agreement with the
extrapolation of the power-law seen at other bands in later epochs.

\section{Observations and analysis}

The observations were obtained in the K$^{\prime}$-band (Wainscoat and
Cowie 1992) with the 3.5-m telescope (equipped with Omega) at the
German-Spanish Calar Alto observatory.  Omega is a near-IR detector with a
1024 $\times$ 1024 pixel HgCdTe array. The image scale is
0.4$^{\prime\prime}$/pixel giving a field of view of 6.8 $\times$ 6.8
arcminutes (Bizenberger et al. 1998). Each image consists of 20 different
frames forming a mosaic on the sky in order to avoid problems due to bad
pixels and cosmic rays. Each frame is the co-addition of 10 images
(3-second exposure time each), see Table \ref{table1} for details.  The two
sets of mosaics were started $3.24$ and $4.77$ hours after the GRB.
Unfortunately no observations could be obtained on Dec 16 (the day after
our measurements) due to bad weather thus preventing a prompt
identification of the near-IR counterpart.

\begin{table}[t]
\begin{center}
\caption{Log of Calar Alto observations for GRB 971214.}
  \begin{tabular}{lccr}
    \label{table1}
   Date           & Number    & Total exposure & K$^{\prime}$ limiting \\ 
   of observations& of frames & time (s)       & magnitude             \\ 
   \hline 
   Dec 15.1076-15.1232 & 20 & 600 & 19.8 \\ 
   Dec 15.1716-15.1875 & 20 & 600 & 19.5 \\ 
   \hline
\end{tabular} 
\end{center}
\end{table}

\begin{figure}[t]
   \centering \resizebox{\hsize}{!}{\includegraphics{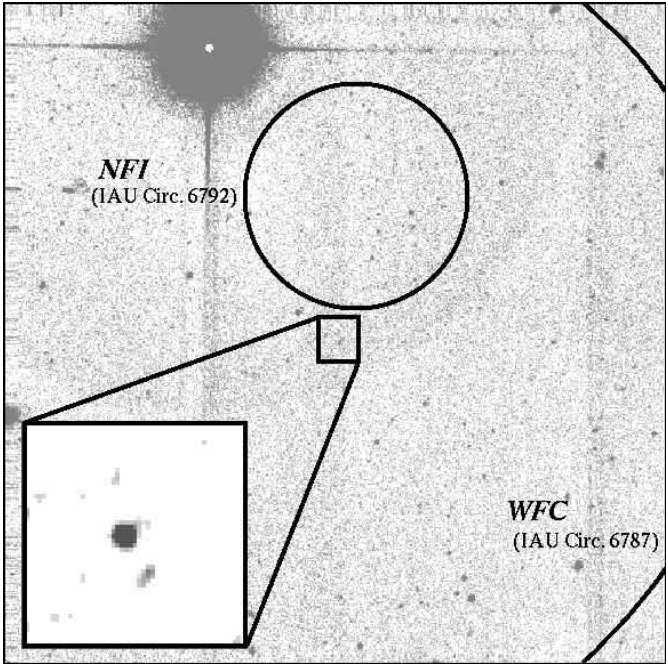}}
  \caption{GRB971214  in the K$^{\prime}$-band as seen by the 3.5-m 
    telescope (+ Omega) at Calar Alto. The image has been obtained
    co-adding the observations taken 3.5 and 5 hours after the gamma-ray
    event. The bright star in the upper left is SAO 15663. The field of
    view is $5.9^{\prime} \times 5.9^{\prime}$, containing $\sim$75\% of
    the WFC error box (the intersections of the WFC error box with the
    image are shown in the right upper and lower corners). The circle
    totally contained in the field of view represents the NFI error box.
    The position of the GRB counterpart, inside the small square, is only
    marginally consistent with the NFI error box.  A blow-up containing the
    GRB counterpart is shown as an inset at the bottom left corner. North
    is at the top and east to the left.}
  \label{fig1}
\end{figure}

Images were calibrated using the UKIRT faint standard FS17 (Casali and
Hawarden 1992), which was imaged immediately after the first GRB mosaic.
Since a K-standard star was used in the calibration, the value of
K-K$^{\prime}$ measured for the GRB depends upon the difference in spectral
shape between the standard star and the GRB transient. Since the transient
is fairly blue (optical spectral index $\sim$ 0) like the standard star,
then K-K$^{\prime} \sim$0 also for the transient. In fact, the maximum
error in the value of K-K$^{\prime}$ is $< 0.05$ magnitudes (Wainscoat and
Cowie 1992), i.e.  not significant compared to the other photometric
errors. The optical transient reported by Halpern et al. (1997) is clearly
detected in both images, $7.5\sigma$ and $5.1\sigma$, respectively, over
the limit (see Fig.~\ref{fig1}). The short time span between both
observations ($\sim$ 1.5 hours) makes it difficult the calculation of the
slope of the light curve.  ~~Photometry of the object~~yields
K$^{\prime}$=$18.03\pm0.18$ for the first observation and
K$^{\prime}$=$18.00\pm0.22$ for the second one.  Therefore the brightness
difference between $\sim$3.5 and $\sim$5 hours after the burst is
$\Delta$K$^{\prime}=-0.03\pm0.28$ magnitudes.  Astrometry~~of~~the~~near-IR
counterpart gives $\alpha= 11^{h} 56^{m} 26.3^{s} \pm0.1^{s}, \delta=
65^{\circ} 12^{\prime} 00^{\prime\prime} \pm 1^{\prime\prime}$ (equinox
2000.0), which is consistent with the optical position reported by Halpern
et al.  (1997).

\section{Discussion}
Prompt follow up observations are extremely important in order to monitor
the GRB afterglow at early stages, which enables to test the validity of
the different afterglow models. In fact, we know now that afterglows are
not always remnants of the initial burst (see Piro et al. 1998).
Observations performed shortly after the GRBs, allowed to measure the delay
between the gamma-ray and the optical emissions in GRB 970228 and GRB
970508, for which the optical maxima were reached 0.7 and 2 days after the
gamma-ray event, respectively.  This appears not to be the case for GRB
971214, for which no optical maximum was detected according to optical
observations started only $\sim$12 hours after the gamma-ray emission
(Halpern et al.  1997).  Therefore the analysis of the K$^{\prime}$-band
data presented in this study, collected $\sim$3.5 and $\sim$5 hours after
the event, is crucial in order to determine whether a delayed emission, at
other wavelengths, following GRB 971214 was present.

\subsection{Multiwavelength spectrum of GRB 971214}
\begin{figure}[t]
  \includegraphics[height=8.8cm,angle=-90]{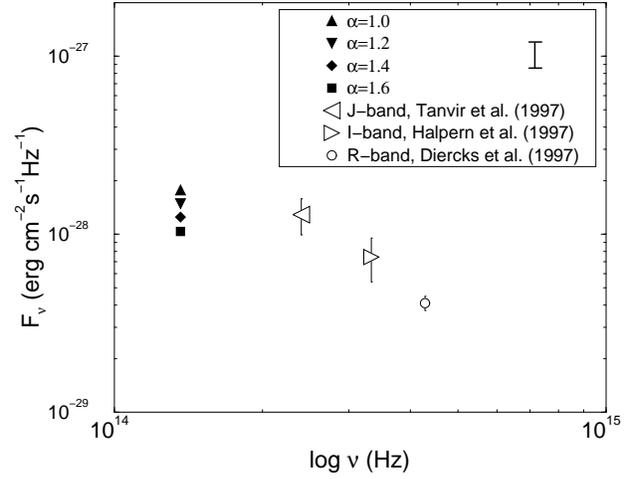}
  \caption{A broad-band spectrum of the GRB 971214 afterglow on December
    15.44-15.51. The points represent measurements in K$^{\prime}$ (solid
    symbols), J, I and R. The values of the K$^{\prime}$-band fluxes have
    been obtained extrapolating the magnitude measured 5 hours after the
    gamma-ray event, for four power-law decay indices $\alpha_{\rm
      K^{\prime}}$ in the range 1.0-1.6. For clarity the error bar in the
    K$^{\prime}$-band is only shown in the upper right corner.}
 \label{fig2}
\end{figure}

Our observations in the K$^{\prime}$-band complete the measurements for the
GRB 971214 afterglow at other wavelengths (J, I, R and X-rays) and are
almost simultaneous to the observation performed with the NFI on board
BeppoSAX, $\sim$6.5 hours after the burst (Antonelli et al.  1997). This
fact enables us to calculate the near-IR to X-ray (2-10 KeV) flux ratio
5-6.5 hours after the GRB, $F_{IR({\rm K}^{\prime})}/F_{X({\rm 2-10~KeV})}
\sim 4.5 \times 10^{-2}$. Taking into account the measurements performed at
other wavelengths on Dec 15.44-15.51, and extrapolating our second
observation with power-law decays ranging from $\alpha_{\rm
  K^{\prime}}=1.0$ to $\alpha_{\rm K^{\prime}}=1.6$, the measured rough
broad-band spectrum (IR-optical) of the GRB 971214 afterglow can be
obtained (see Fig.~\ref{fig2}). As it can be seen in this figure, the shape
of the measured flux density distribution $F_{\nu}$ vs $\log \nu$ depends
on the power-law index assumed for the ${\rm K}^{\prime}$ light curve.
Thus, if $\alpha_{\rm K^{\prime}} > 1.1$ the multiwavelength spectrum of
the GRB 971214 afterglow would show a maximum in the near-IR. This possible
maximum around the J-band has not been detected in previous multiwavelength
spectra of GRB 971214 due to the lack of measurements in the ${\rm
  K^{\prime}}$-band (Reichart 1998).

\subsection{Study of the light curve}
According to the fireball models, the afterglow radiation will shift
progressively to lower frequencies and the corresponding timescales will
lengthen (Katz and Piran 1997). Therefore, the K$^{\prime}$-band
measurements impose, over the available data, the most stringent limits to
a possible maximum in the light curve of the GRB 971214 afterglow. Three
possible light curve shapes can fit our data: a rising light curve (case 1)
similar to that detected for GRB 970228 (Guarnieri et al.  1997, Pedichini
et al. 1997) and GRB 970508 (Pedersen et al.  1998, Galama et al.  1998), a
plateau phase (case 2), as it was seen in GRB 970508 between $\sim$ 4 and
$\sim$ 24 hours after the burst (Castro-Tirado et al.  1998), or a
power-law decay (case 3) as it was later reported for both GRB 970228
(Galama et al.  1997) and GRB 970508 (Sokolov et al.  1998).

\subsubsection{Non fading light curve}
In case of a rising light curve (case 1) or a plateau (case 2) the light
curve could not extend, following the same trend, until Dec 15.54, giving
the upper limit of K $>$ 18.5 imposed by Garcia et al.  (1997) (see
Fig.~\ref{fig3}).

Even an increasing curve until a time near Dec 15.54 followed by a very
sharp bending over would be unrealistic. We have estimated the acceptable
time for this turning in the following way: first we have assumed that the
light curve displayed a plateau phase (at K$^{\prime}=18.0$) from Dec 15.18
onwards; then we have constructed a curve with a power-law decline index
$\alpha_{\rm K^{\prime}}=1.6$ that matches the Garcia et al. (1997)
measurement on Dec 15.54. The intersection between the constant light curve
(initial phase) at K$^{\prime}$=18.0 and the power-law fading light curve
provides an upper limit for the time of the turnover point, T$_{p}$ (see
Fig.~\ref{fig3}). On the other hand, if we assume an increasing light curve
crossing our two data points at Dec 15.12 and Dec 15.18, the intersection
with the power-law $\alpha_{\rm K^{\prime}}=1.6$ would move slightly
backwards.  Lower values of $\alpha_{\rm K^{\prime}}$ would also give lower
values of T$_{p}$ in all cases. We have not considered larger (unrealistic)
values of $\alpha_{\rm K^{\prime}}$. So, we conclude that the possible
maximum or turning point took place T$_{p}$=10.2 hours after the gamma-ray
event, at the latest.

\subsubsection{Fading light curve}

\begin{figure}[t]
   \includegraphics[height=8.8cm,angle=-90]{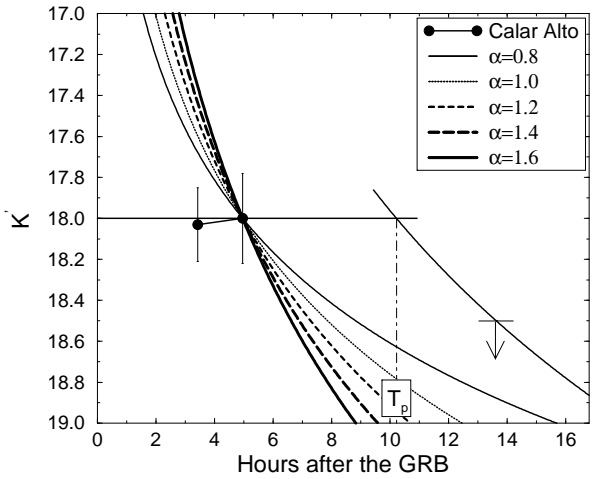}
  \caption{The solid circles represent the results of the observations 
    carried out at Calar Alto 3.5 and 5 hours after the burst. The arrow
    shows the upper limit found by Garcia et al. (1997). The curves
    represent the power-law decays for different exponents, ranging from
    $\alpha_{\rm K^{\prime}}=0.8$, to $\alpha_{\rm K^{\prime}}=1.6$. The
    upper limit does not impose any constraint unless $\alpha_{\rm
      K^{\prime}} < 0.45$.  The intersection of the horizontal line with
    the power-law curve that goes through the Garcia et al. measurement
    provides an upper limit for the turning point time, T$_{p}$=10.2
    hours.}
 \label{fig3}
\end{figure}

If we assumed a single-fading light curve (case 3) with a power-law decline
in the near-IR with index $\alpha_{\rm K^{\prime}}=1.2$ (as used by Waxman
(1997) and similar to the other two optical counterparts) a variation
$\Delta$K$^{\prime}$ between our images taken $\sim$1.5 hours apart of
$\Delta$K$^{\prime}=0.484$ magnitudes would be expected.  However, our data
imply a magnitude difference $\Delta$K$^{\prime}=-0.03\pm 0.28$, which is
$1.7\sigma$ from the above mentioned prediction derived from the
$\alpha_{\rm K^{\prime}}=1.2$ power-law (see Fig.~\ref{fig3}). If the
assumed power-law index were $\alpha_{\rm K^{\prime}}=1.4$ the rejection
level would be $2.0\sigma$, being necessary a power-law index $\alpha_{\rm
  K^{\prime}}=2.1$ (too unrealistic) in order to find a disagreement at a
$3.0\sigma$ level between our points and the prediction of a power-law
decay. The power-law light curve connecting our second measurement
$\sim$5.0 hours after the gamma-ray event and the upper limit imposed by
Garcia et al. (1997), would have an index $\alpha_{\rm K^{\prime}}=0.45$.
Therefore, power-law decays with indices $\alpha_{\rm K^{\prime}} \leq
0.45$ are ruled out, because they imply a magnitude $K \leq 18.5$ on Dec
15.54, which would be above the reported upper limit.

\section{Conclusions}
We have detected the GRB 971214 near-IR counterpart $\sim$3.5 hours and
$\sim$5 hours after the gamma-ray event which enables to conclude that:

i) a magnitude difference $\Delta$K${^{\prime}}=-0.03\pm 0.28$ is derived
from our measurements, whereas $\Delta$K${^{\prime}}=0.464$ would be
expected assuming a power-law decay with index $\alpha_{\rm
  K^{\prime}}=1.2$ (similar to the one observed at optical wavelengths).
This implies a deviation of $1.7\sigma$. If the assumed power-law index
$\alpha_{\rm K^{\prime}}$ were 1.4, then the rejection level would be
$2.0\sigma$.  Thus, our measurements suggest a rising or a flat light curve
segment with a duration $ 1.5 \leq T \le 6.7 $ hours (between 3.5 and 10.2
hours after the burst).  This conclusion must be taken with care since the
above-mentioned rejection levels are not stringent enough to assure the
result with total confidence;

ii) the power-law decline in the near-IR should be steeper than the one
given by $\alpha_{\rm K^{\prime}}=0.45$;

iii) for the observations carried out on Dec 15.44-15.51, there is a change
in the slope of the measured energy distribution at some wavelength between
the J and K$^{\prime}$ bands (possibly around H).

\section{Acknowledgments}
We thank A. Fruchter, B. Montesinos and E. Pian for fruitful conversations.
This work has been partially supported by Spanish CICYT grant
ESP95-0389-C02-02.

\end{document}